\documentclass[twocolumn,10pt]{IEEEtran}
\usepackage{ifpdf, flushend,subfigure}
\usepackage{graphicx}
\usepackage{epstopdf}
\graphicspath{{../}}
\usepackage{multirow}
\usepackage{float}
\usepackage[cmex10]{amsmath}
\usepackage {amssymb}
\usepackage{array}
\usepackage{mdwmath}
\usepackage{mdwtab}
\usepackage{eqparbox}
\usepackage{url}
\usepackage{nomencl}
\usepackage{cite}
\usepackage{algorithm}
\usepackage{algorithmic}
\usepackage{makeidx}
\usepackage{ifthen}
\usepackage{subfigure}
\usepackage{caption}
\usepackage{pdflscape}
\usepackage{hyperref}
\usepackage{xcolor}
\def\bz{\mathbf{z}}
\def\mbi{\mathbb{I}}

\def\rev#1{{\color{black}#1}} 

\newcommand{\bbeta}{{\boldsymbol{\beta}}}
\makeatletter
\def\munderbar#1{\underline{\sbox\tw@{$#1$}\dp\tw@\z@\box\tw@}}
\makeatother

\usepackage{amsmath,mathtools, amssymb, bbm, xspace}
\usepackage{booktabs,ragged2e}
\usepackage{multirow}
\usepackage[flushleft]{threeparttable}

\usepackage{graphicx}

\def\det{\mathop{\hbox{\rm det}}}
\def\diag{\mathop{\hbox{\rm diag}}}

\def\spose#1{\hbox to 0pt{#1\hss}}

\def\text #1{\hbox{\quad#1\quad}}

\def\nthinsp{\mskip -2   mu}

\def\G{_{\scriptscriptstyle G}}

\def\R{_{\scriptscriptstyle R}}

\def\superstar{^{\raise 0.5pt\hbox{$\nthinsp *$}}}
\def\SUPERSTAR{^{\raise 0.5pt\hbox{$*$}}}

\def\lamstarT {\lambda^{\raise 0.5pt\hbox{$\nthinsp *$}T}}

\def\hbar{\skew{4.2}\bar h}

		\def\bkE{{\rm I\kern-.17em E}}
		\def\bk1{{\rm 1\kern-.17em l}}
		\def\bkD{{\rm I\kern-.17em D}}
		\def\bkR{{\rm I\kern-.17em R}}
		\def\bkP{{\rm I\kern-.17em P}}
		\def\bkY{{\bf \kern-.17em Y}}
		\def\bkZ{{\bf \kern-.17em Z}}

		\def\bc{\begin{center}}
		\def\be{\begin{enumerate}}
		\def\bi{\begin{itemize}}
		\def\ec{\end{center}}
		\def\ee{\end{enumerate}}
		\def\ei{\end{itemize}}
		\def\es{\end{small}}
		\def\eS{\end{slide}}

	\def\cp2problem#1#2#3#4{\fbox
		 {\begin{tabular*}{0.9\textwidth}
			{@{}l@{\extracolsep{\fill}}l@{\extracolsep{6pt}}l@{\extracolsep{\fill}}c@{}}
				#1 & & $#4 $ 
			\end{tabular*}}}

		\def\bkE{{\rm I\kern-.17em E}}
		\def\bk1{{\rm 1\kern-.17em l}}
		\def\bkD{{\rm I\kern-.17em D}}
		\def\bkR{{\rm I\kern-.17em R}}
		\def\bkP{{\rm I\kern-.17em P}}
		
		\def\bkZ{{\bf{Z}}}

\newcommand {\beeq}[1]{\begin{equation}\label{#1}}
\newcommand {\eeeq}{\end{equation}}
\newcommand {\bea}{\begin{eqnarray}}
\newcommand {\eea}{\end{eqnarray}}

\def\texitem#1{\par\smallskip\noindent\hangindent 25pt
               \hbox to 25pt {\hss #1 ~}\ignorespaces}

\newcommand{\argmin}{\operatornamewithlimits{argmin}}

\newcommand{\beq}{\begin{equation}}
\newcommand{\eeq}{\end{equation}}
\newcommand{\beqn}{\begin{eqnarray}}
\newcommand{\eeqn}{\end{eqnarray}}
\newcommand{\beqno}{\begin{eqnarray*}}
\newcommand{\eeqno}{\end{eqnarray*}}
\newcommand{\bma}{\begin{displaymath}}
\newcommand{\ema}{\end{displaymath}}
\newcommand{\bnu}{\begin{enumerate}}
\newcommand{\enu}{\end{enumerate}}
\newcommand{\bce}{\begin{center}}
\newcommand{\ece}{\end{center}}
\newcommand{\btb}{\begin{tabular}}
\newcommand{\etb}{\end{tabular}}

\def\R{{\mathbb{R}}}
\def\G{{\mathbb{G}}}

\def\E{{\mathsf{E}}}

\def\bx{{\mathbf{x}}}
\def\by{{\mathbf{y}}}
\def\bz{{\mathbf{z}}}
\def\bu{{\mathbf{u}}}
\def\bv{{\mathbf{v}}}
\def\bL{{\mathbf{L}}}
\def\bF{{\mathbf{F}}}
\def\b1{{\mathbf{1}}}
\def\cNinik{{\mathcal{N}^{\rm in}_{ik}}}
\def\cNoutik{{\mathcal{N}^{\rm out}_{ik}}}
\def\cNini{{\mathcal{N}^{\rm in}_{i}}}
\def\cNouti{{\mathcal{N}^{\rm out}_{i}}}
\def\diag{{\rm diag}}

\usepackage[normalem]{ulem} 

\newtheorem{theorem}{Theorem}

\newtheorem{proposition}{Proposition}
\newtheorem{lemma}{Lemma}
\newtheorem{remark}{Remark}

\newtheorem{assumption}{Assumption}

\newcommand{\bg}{{\mathbf{g}}}

\newcommand{\g}{{\mathbf{g}}}

\newcommand{\Diag}{{\mathrm{Diag}}}

\newcommand{\ra}{{r_\mathrm{a}}}

\newcommand{\one}{{\mathbf{1}}}
\newcommand{\zero}{{\mathbf{0}}}

\newcommand{\T}{{\mathsf{T}}}

\definecolor{myBlue}{rgb}{0.80,0.85,1.00}
\definecolor{myYellow}{rgb}{0.951,1.000,0.547}

\def\la{{\langle}}
\def\ra{{\rangle}}

\def\bit{\begin{itemize}}
\def\eit{\end{itemize}}
\def\BEAS{\begin{eqnarray*}}
\def\EEAS{\end{eqnarray*}}

\def\re{{\mathbb R}}

\def\a{\alpha}
\def\b{\beta}

\def\g{\gamma}

\def\argmin{\mathop{\rm argmin}}

\begin{document}

\title{Geometric Convergence of Distributed Heavy-Ball Nash Equilibrium Algorithm over Time-Varying Digraphs with Unconstrained Actions}

\author{Duong Thuy Anh Nguyen, \IEEEmembership{Student Member, IEEE},
Duong Tung Nguyen, \IEEEmembership{Member, IEEE}, \\and Angelia Nedi\'c, \IEEEmembership{Member, IEEE}
\thanks{The authors are with the School of Electrical, Computer and Energy Engineering, Arizona State University, Tempe, AZ, United States. 
Email: \{dtnguy52,~duongnt,~Angelia.Nedich\}@asu.edu. This material is based in part upon work supported by the NSF award CCF-2106336.
\textit{Corresponding Author}: Duong Thuy Anh Nguyen.}
}

\maketitle

\begin{abstract} 
This paper presents a new distributed algorithm that leverages heavy-ball momentum and a consensus-based gradient method to 
find a Nash equilibrium (NE) in a class of non-cooperative convex games with unconstrained action sets. In this approach, each agent in the game has access to its own smooth local cost function and can exchange information with its neighbors over a communication network. 
\rev{The main novelty of our work is the incorporation of heavy-ball momentum in the context of non-cooperative games that operate on fully-decentralized, directed, and time-varying communication graphs, while also accommodating non-identical step-sizes and momentum parameters. Overcoming technical challenges arising from the dynamic and asymmetric nature of mixing matrices and the presence of an additional momentum term, we provide a rigorous proof of the geometric convergence to the NE.} Moreover, we establish explicit bounds for the step-size values and momentum parameters based on the characteristics of the cost functions, mixing matrices, and graph connectivity structures. We perform numerical simulations on a Nash-Cournot game to demonstrate accelerated convergence of the proposed algorithm compared to that of the existing methods.
\end{abstract}

\printnomenclature
\allowdisplaybreaks

\section{Introduction} \label{intro}
Nash equilibrium (NE) computation is essential for examining decision-making and strategic behavior in multi-agent systems, especially in non-cooperative games. These games have applications in numerous engineering domains, including electricity markets, power systems, flow control, and crowdsourcing~\cite{ BasharSG,wiopt23}.
In non-cooperative games, each agent has their own goals and seeks to maximize their payoff, without coordinating with other agents. Classical complete information game theory typically 
employs best-response or gradient-based methods to find an NE, but these approaches
necessitate each agent having complete information regarding the actions of its competitors~\cite{Yi2019,Belgioioso2018}, which can be unrealistic in many practical engineering systems.

Extensive research has 
been conducted to develop efficient distributed methods for seeking NE \rev{in the settings with partial information available to the agents}. These methods are primarily built upon projected gradient and consensus dynamics approaches, and they have been studied in both continuous-time~\cite{Gadjov2019} and discrete-time~\cite{Koshal2016,Tatiana2020} domains. 
Early works only consider time-invariant undirected networks,
such as for example~\cite{SALEHISADAGHIANI201927}, which develops an algorithm within the framework of an inexact-ADMM.
The accelerated gradient play algorithm (Acc-GRANE) presented in~\cite{Tatiana2018} is based on the strong monotonicity of an augmented mapping and is applicable to a subclass of games. 
Reference~\cite{Tatarenko2021} 
extends the applicability of the Acc-GRANE algorithm to a broader class of games by assuming the restricted strong monotonicity of the augmented mapping.
\rev{Based on the contraction properties of doubly stochastic matrices,
reference~\cite{Tatiana2020} develops a distributed gradient method}  whose convergence properties do not depend on an augmented mapping.

There has been a growing interest in studying NE computation in communication networks with switching topologies. Early work~\cite{Koshal2016} focuses on aggregative games over 
undirected graphs, while~\cite{Grammatico2021} extends the study to games with coupling constraints. \rev{Reference~\cite{Farzad2019} proposes an asynchronous gossip algorithm for a directed graph, assuming each agent can update all estimates of the agents who interfere with its cost function.} In~\cite{Bianchi_2021}, a projected pseudo-gradient-based algorithm is proposed for time-varying directed graphs that are weight-balanced, while~\cite{Bianchi2020NashES} relaxes the weight-balancedness assumption for a static directed graph, assuming complete knowledge of the underlying communication graph. \rev{In~\cite{nguyen2022distributed}, these assumptions are further relaxed, and a condition is established to address the loss of monotonicity with weighted norms, which is a commonly encountered issue. In~\cite{GadjovTCSN2023}, a robust algorithm is proposed that utilizes an observation graph allowing for direct observation of actions and, thus, making it immune to tampering by adversarial agents.}

Most of the above mentioned works deal with gradient-based methods. The heavy-ball method, introduced in~\cite{Polyak}, has been widely used as an acceleration technique for gradient-based methods to achieve faster convergence~\cite{Nguyen2023AccAB}. In the context of aggregative games, the heavy-ball method has been employed in algorithms for both undirected~\cite{Song2020} and directed~\cite{Fang2022} graphs, achieving convergence for diminishing step-sizes. \rev{It has also been employed in semi-decentralized communication architectures, where 
a central coordinator collects and transmits aggregative signals to the agents in the system~\cite{BelgioiosoTAC2023}. 
Reference~\cite{BIANCHI2022110080} considers games over undirected graphs with affine coupling constraints and introduces acceleration schemes for the proximal-point algorithm, such as (alternated) inertia and overrelaxation.
}

\textbf{Contributions.} Motivated by the potential benefits of the heavy-ball method in accelerating convergence, this paper presents a novel distributed, discrete-time algorithm for NE seeking by integrating the heavy-ball momentum and consensus-based gradient method. \rev{The proposed algorithm is designed to work on a general sequence of time-varying directed graphs, without requiring any coordination among agents for the weights. It also allows for non-identical step-sizes and momentum parameters.
The incorporation of agent-based heavy-ball momentum terms introduces additional technical complexity in the convergence analysis of the algorithm. The main challenge is in ensuring the contraction properties of a recurrence relation due to the presence of the additional momentum terms, especially when dealing with time-varying asymmetric mixing matrices. Prior heavy-ball algorithms, such as those presented in~\cite{Song2020} and~\cite{Fang2022} for aggregative games, utilize diminishing step-sizes to demonstrate that the state difference resulting from the additional momentum term converges to zero. Unfortunately, this approach is not applicable to this work since 
our algorithm employs a constant step-size rule. To overcome this technical challenge, we analyze three distinct quantities: the consensus error, the NE gap, and consecutive states difference. By establishing a composite relation for the vector comprising these quantities, we rigorously prove the linear convergence of the algorithm to the NE under strong monotonicity and Lipschitz continuity assumptions. We also provide explicit bounds for the constant step-size values and momentum terms based on the properties of the cost functions, mixing matrices, and graph connectivity structures.}


This paper is structured as follows. Section~\ref{sec:formu} presents the problem formulation. In Section~\ref{sec:algo}, the distributed algorithm is introduced. Section~\ref{sec:conv_results} provides convergence analysis and Section~\ref{sec:simulation} numerically evaluates the performance of the proposed algorithm. Section~\ref{sec:conc} concludes with key points.

\textbf{Notations.} 
All vectors are column vectors unless otherwise stated. We write $u^\T$ for the transpose of a vector $u\in \re^n$. We use $\zero$ and $\one$ to denote the vector with all entries equal to $0$ and $1$, respectively.  The $i$-th entry of a vector $u$ is denoted by $u_i$, while it is denoted by $[u_k]_i$ for a time-varying vector $u_k$. We denote $\min(u)=\min_i u_i$ and $\max(u)=\max_i u_i$.  \rev{Given a vector $u$, $\Diag(u)$ denotes the diagonal matrix whose diagonal entries correspond to the entries of $u$.} A nonnegative vector is called stochastic if its entries \rev{sum up to $1$}. 

We use $A_{ij}$ to denote the $ij$-th entry of a matrix $A$, and $[A_k]_{ij}$ when the matrix is time-dependent. The notation $A\le B$ is used when $A_{ij}\le B_{ij}$ for all $i, j$, where $A$ and $B$ are matrices of the same dimension. A matrix $A$ is nonnegative if all its entries are nonnegative and ${\min}^{+}(A)$ denotes the smallest positive entry of $A$. A nonnegative matrix $A\in\mathbb{R}^{n\times n}$ is row-stochastic if $A\mathbf{1}=\mathbf{1}$. A \textit{consensual} matrix has equal row vectors. The identity matrix is denoted by $\mbi$. Given a vector $\pi\in\re^m$ with positive entries, we denote
\vspace{-0.2cm}
\begin{center}
    $\la \bu,\bv\ra_{\pi}=\sum_{i=1}^m \pi_i\la u_i,v_i \ra $ and $\|\bu\|_{\pi}=\sqrt{\sum_{i=1}^m \pi_i\|u_i\|^2},$
\end{center}
where \rev{$\bu\!:=\![u_1,\ldots,u_m]^\T, \bv \!:=\![v_1,\ldots,v_m]^\T \!\!\in\! \re^{m\times n}$}, and $u_i,v_i\!\in\!\re^n$. When $\pi = \one$, we write $\la \bu,\bv\ra$ and $\|\bu\|$. We have
\begin{align}\label{eq-NormIneq}   \tfrac{1}{\sqrt{\max(\pi)}}\|\bu\|_{\pi}
\le \|\bu\| 
  \le \tfrac{1}{\sqrt{\min(\pi)}}\|\bu\|_{\pi},
\end{align}
and $\la \bu,\bv\ra_{\pi}    \le\|\bu\|_{\pi}\|\bv\|_{\pi}$ (Cauchy–Schwarz inequality).


We let $[m]=\{1,\ldots,m\}$ for an integer $m\ge 1$. 
Given a directed graph $\G=([m],\E)$, specified by the set of edges $\E\subseteq [m]\times[m]$ of ordered pairs of nodes, the in-neighbor and out-neighbor set for every agent $i$ are defined, as follows:
\[\cNini=\{j\!\in\![m]|(j,i)\!\in\!\E\} \!\!\!\text{and}\!\!\! \cNouti=\{\ell\!\in\![m]|(i,\ell)\!\in\!\E\}.\] 
A directed graph $\G$ is {\it strongly connected} if there is a directed path from any node to all other nodes in $\G$. 
We use $\mathsf{D}(\G)$ and $\mathsf{K}(\G)$ to denote the diameter and the maximal edge-utility of a strongly connected directed graph $\G$, respectively, as defined in Definition 2.1 and Definition 2.2 in \cite{Angelia2022AB}.

\section{Problem Formulation} \label{sec:formu}
We study a non-cooperative game with $m$ agents, where each agent has an unconstrained action set $X_i = \re^{n_i}$, for $i\in[m]$. Each agent $i$ has a cost function $J_i(x_i,x_{-i})$ that depends on its own action $x_i \in X_i$ and the joint action of all other agents except itself, $x_{-i} \in X_{-i}=\re^{n-n_i}$. The joint action vector of all agents has size $n = \sum_{i=1}^m n_i$ and belongs to the joint action set $X = X_1\times\cdots\times X_m = \re^n$. We assume that the cost function $J_i(x_i,x_{-i})$ is continuously differentiable in $x_i$ for any fixed $x_{-i}\in X_{-i}$, for all $i\in [m]$.

Denote the game by $\Gamma=([m],\{J_i\},\{X_i\})$.
A vector $x^*=[x_1^*,\ldots,x_m^*]^\T\in X$ is a NE to the game $\Gamma$ if, for every agent $i\in[m]$, the condition below is satisfied:
\beqn
J_i(x_i^*,x_{-i}^*)\le J_i(x_i,x_{-i}^*),  \quad\hbox{for all } x_i\in X_i.
\eeqn
We define the game mapping $F(x):X\to\re^n$ as 
\begin{align}\label{eq:gamemapping}
F(x)\triangleq\left[\nabla_1 J_1(x_1,x_{-1}), \ldots, \nabla_m J_m(x_m,x_{-m})\right]^\T,
 \end{align}
where $\nabla_i J_i(x_i,x_{-i})=\nabla_{x_i} J_i(x_i,x_{-i})$ for all $i\in[m]$.

We make the following assumptions.

\begin{assumption}\label{assum:map_monotone}
The game mapping $F(x)$ is strongly monotone on $X$ with the constant $\mu>0$.
\end{assumption}
\begin{assumption}\label{assum:lip}
Consider the game $\Gamma=([m],\{J_i\},\{X_i\})$, assume for all $i \in [m]$:\\
(a) The mapping $\nabla_i J_i(x_i,\cdot)$ is Lipschitz continuous on
$X_{-i}$ for every fixed $x_i\in X_i$ with a uniform constant $L_{-i}>0$.\\
(b) The mapping $\nabla_i J_i(\cdot,x_{-i})$ is Lipschitz continuous on $X_i$ for every fixed $x_{-i}\in X_{-i}$ with a uniform constant $L_i>0$.
\end{assumption}
\begin{remark}
Assumption~\ref{assum:map_monotone} implies strong convexity of each cost function $J_i(x_i,x_{-i})$ on $X_i$ for every $x_{-i}\in X_{-i}$ with the constant $\mu$, as noted in Remark 1 of \cite{Tatiana2018}. The existence and uniqueness of a NE for the game $\Gamma=([m],{J_i},{X_i})$ is also guaranteed by Assumption~\ref{assum:map_monotone}. This result is established in Theorem 2.3.3 of \cite{FacchineiPang}. Moreover, as $X_i=\re^{n_i}$ for all $i\in [m]$, $x^*\in X$ is the NE if and only if $F(x^*)=\zero$.
\end{remark}

\section{Distributed Nash Equilibrium Seeking} \label{sec:algo} 
Consider a game $\Gamma=([m],{J_i},{X_i})$ where agents interact through a sequence of time-varying communication graphs $\G_k=([m],\E_k)$ at time $k$. Each link $(j,i)$ in $\E_k$ indicates that agent $i$ receives information from agent $j$. The graph is assumed to have self-loops for every node in each $\G_k$, ensuring that $\cNinik$ and $\cNoutik$ always contain agent $i$. The details are outlined in the following assumption.

\begin{assumption}\label{asum-graphs}
Each graph $\G_k=([m],\E_k)$ is strongly connected and has a self-loop at every node $i\in[m]$.
\end{assumption}

\begin{remark}\label{rem-graphs}
Assumption~\ref{asum-graphs} can be relaxed by considering $B$-strongly-connected graph sequence, i.e., \rev{when an integer {$B\ge 1$} exists} such that the graph with edge set 
$\E^B_k=\bigcup_{i=kB}^{(k+1)B-1}\E_i$
is strongly connected for every $k\ge 0$.
\end{remark}

To deal with the partial information available to agents, each agent $i$ maintains a local variable $z_i^{k}=(z_{i1}^k,\ldots,z_{im}^k)^\T\in\re^n$, where $z_{ij}^k$ is agent $i$'s estimate of the decision $x_j^k$ for agent $j\ne i$, while $z_{ii}^k=x_i^k$. The estimate of agent $j$ without the $i$-th block-component is defined as
\[z_{j,-i}^k=(z_{j1}^k,\ldots,z_{j,i-1}^k, z_{j,i+1}^k,\ldots,z_{jm}^k)^\T\in\re^{n-n_i}.\]

Given the constraints on agents' access to others' actions in game $\Gamma$, we propose a fully-distributed algorithm that respects the information access as dictated by the communication graphs $\G_k$. The approach, outlined in Algorithm 1, incorporates a gradient method with heavy-ball momentum. At each time $k$, each agent $i$ sends its estimate $z_i^k$ to its out-neighbors $\ell\in\cNoutik$ and receives estimates $z_j^k$ from its in-neighbors $j\in\cNinik$. Agent $i$ then updates its own action $x_i^{k+1}$ and local estimate $z_i^{k+1}$ using the received information.

\begin{table}[t!]
\centering \normalsize
\vspace{0.25cm}
    \begin{tabular}{l}
    \hline
    \multicolumn{1}{c}{\textbf{Algorithm 1: DNE-HB}}\\
    \hline
    Every agent $i\in[m]$ selects a local stepsize $\a_i>0$, a\\ local momentum parameter $\beta_i>0$ and initializes with\\ arbitrary initial vectors $z_{i,-i}^0\in\re^{n-n_i},x_i^{0}, x_i^{-1}\in\re^{n_i}$.\\
    \textbf{for} $k=0,1,\ldots,$ every agent $i\in[m]$ does the following:\!\!\\
    \emph{  } Receives $z_j^k$ from in-neighbors $j\in\cNinik$;\\
    \emph{  } Sends $z_i^k$ to  out-neighbors $\ell\in\cNoutik$;\\
    \emph{ } Chooses the weights $[W_k]_{ij},j\in[m]$;\\
    \emph{ } Updates the action $x_i^{k+1}$ and estimates $z_{i,-i}^{k+1}$ by \\
    \emph{~~~~} $x_{i}^{k+1}=\sum_{l=1}^m [W_k]_{il}z_{li}^k-\a_i\nabla_i J_i\!\left(\sum_{j=1}^m [W_k]_{ij}z_{j}^k\right)$\\
    \emph{~~~~}$\qquad ~~~~ +\beta_i(x_i^k-x_i^{k-1})$;~ $z_{i,i}^{k+1}=x_{i}^{k+1}$;\\
    \emph{~~~~} $z_{i,-i}^{k+1}=\sum_{j=1}^m [W_k]_{ij}z_{j,-i}^k+\!\beta_i(z_{i,-i}^k\!-z_{i,-i}^{k-1})$;\\
    \textbf{end for}\\
    \hline
    \end{tabular}
    \vspace{-0.5cm}
\end{table}
We make the following assumption on the matrices $W_k$.
\begin{assumption} \label{asum-row-stochastic}
For each $k\ge0$, the weight matrix $W_k$ is row-stochastic and compatible with the graph $\G_k$
i.e., 
\begin{align}\label{eq-mat-compat}
\begin{cases}
[W_k]_{ij}>0, \quad\text{when} j\in\cNinik,\\
[W_k]_{ij}=0, \quad\text{otherwise.}
\end{cases}
\end{align}
There exist a scalar $w\!>\!0$ such that $\min^{+}(W_k)\ge w, \forall k\ge 0$.
\end{assumption}


\section{Geometric Convergence of \textbf{DNE-HB}}\label{sec:conv_results}
\subsection{Preliminaries}\label{sec:basicre}
We outline basic results on norm of linear combinations of vectors, graphs, stochastic matrices, and gradient method.

\begin{lemma}[\!\!\cite{nguyen2022distributed}, Corollary~5.2]\label{lem-normlincomb}
Consider a vector collection $\{u_i, \, i\in[n]\}\subset\re^p$, and a scalar collection $\{\g_i,\, i\in[n]\}\subset\re$ of scalars such that $\sum_{i=1}^n \g_i=1$. For all $u\in \re^p$, we have the following relation:
\begin{align*}
    \Bigg\|\!\sum_{i=1}^n \!\g_i u_i \!-\! u \Bigg\|^2 \!\!\!= \!\sum_{i=1}^n \!\g_i \|u_i\!-\!u\|^2 \!-\!\!\sum_{i=1}^n \g_i \Bigg\|u_i \!-\! \Bigg(\sum_{\ell=1}^n \g_\ell u_\ell\Bigg)\!\Bigg \|^2\!\!\!.
\end{align*}
\end{lemma}

    
	
\begin{lemma}[\!\!\cite{nguyen2022distributed}, Lemma 5.4]\label{lemma-TimeVaryPiVectors}
\rev{Under Assumption~\ref{asum-graphs}, when} ${W_k}$ satisfies Assumption~\ref{asum-row-stochastic}, we have for all $k\ge 0$:
\begin{enumerate}
\item[(a)] There exists a sequence $\{\pi_k\}$ of stochastic vectors such that 
$\pi_{k+1}^\T W_k=\pi_k^\T $. 
\item[(b)] The entries of each $\pi_k$ have a uniform lower bound,
i.e.,
$[\pi_k]_i\ge\frac{w^m}{m}$ for all $i\in[m]$ and all $k\ge0$.
\end{enumerate}
\end{lemma}
Using the stochastic vectors $\pi_k$ described in Lemma~\ref{lemma-TimeVaryPiVectors}, we can define an appropriate Lyapunov function for the method.


\begin{lemma}[\!\!\cite{nguyen2022distributed}, Lemma 6.1]\label{lemma-lemma6PushPull}
Let $\G=([m],\E)$ be a strongly connected directed graph, and let $W$ be an $m\times m$ row-stochastic matrix that is compatible with the graph and has positive diagonal entries. Also, let $\pi$ be a stochastic vector and let $\phi$ be a nonnegative vector such that $\phi^\T W=\pi^\T.$
Consider a collection of vectors $z_{1},\ldots,z_{m} \in \re^n$ and consider the vectors $r_i=\sum_{j=1}^m W_{ij}z_j$, for all $i\in[m]$, and let $\hat{z}_{\pi}=\sum_{i=1}^m \pi_{i}z_i$, for all $u\!\in\! \re^n$, we have
\begin{align*}
    \sum_{i=1}^m&\phi_i\left\lVert r_i-u\right\rVert^2
    \le \sum_{j=1}^m\pi_j\|z_j-u\|^2 \\
    &-\dfrac{\min(\phi)\left({\min}^{+}(W)\right)^2}{\max^2(\pi)\mathsf{D}(\G)\mathsf{K}(\G)}\sum_{j=1}^m\pi_j\|z_j-\hat{z}_{\pi}\|^2.
\end{align*}
\end{lemma}

Let $\bz_{i:}$ be the vector in the $i$th row of matrix $\bz\in\re^{m\times n}$. We define a mapping $\bF_\a(\cdot)$ where the $i$th row is given by
\begin{equation}\label{eq-def-bfa}
\!\!\!\![\bF_\a(\bz)]_{i:}\!=\!(\zero^\T_{n_1}\!,\!..., \zero^\T_{n_{i-1}}\!, \a_i(\nabla_i J_i(\bz_{i:}))^\T\!, \zero^\T_{n_{i+1}}\!,\!...,\!\zero^\T_{n_m} ).\!\!\!
\end{equation}

\begin{lemma}[\rev{\!\! \cite{nguyen2022distributed}, \!Lemma 5.6}]\label{lemma-LipschitzbF}
\!Let Assumptions~\ref{assum:lip} hold and $\bL_\a\!\!=\!\!\sqrt{\max\limits_{i\in[m]}\{\!\a_i^2(L_{-i}^2\!+\! L_i^2)\!\}}$.
For any stochastic vector $\pi\!>\!\zero$, 
\begin{align*} 
    \!\!\|\bF_\a{(\bz)}-\bF_\a{(\by))}\|_\pi^2 \le \bL_\a^2\|\bz-\by\|_\pi^2 \!\!\!\!\text{for all}\!\!\!\! \bz,\by\in\re^{m\times n}.
\end{align*}
\end{lemma}

\begin{lemma}[\rev{\!\! \cite{Polyak}}]\label{lem-contraction} For a $\mu$-strongly convex function $f$ with $L$-Lipschitz continuous gradients, at the point $x^*=\argmin_{x} f(x)$, for all $\a$ with $0<\a<2L^{-1}$, we have
\[\|x-x^*-\a \nabla f(x)\|\le q(\a) \|x-x^*\|\quad\hbox{for all $x$},\]
where $q(\a)=\max\{|1-\a\mu|,|1-\a L|\}<1$. 
\end{lemma}

\subsection{Convergence Results}\label{sec:conv}
Consider the sequence of time-varying directed graphs $\G_k=([m],\E_k)$. 
Under assumptions \ref{assum:map_monotone}-\ref{asum-row-stochastic}, we provide a proof demonstrating that the iterate sequence ${x^k}$ generated by \textbf{DNE-HB} exhibit geometric convergence towards the NE.

Let $z_i^{k}=(z_{i1}^k,\ldots,z_{im}^k)^\T\in\re^n$ for all $k\ge0$, and let $\{\pi_k\}$ denote the sequence of stochastic vectors satisfying $\pi_{k+1}^\T W_{k}=\pi_k^\T$, with $\pi_k>\zero$. We define matrices
\begin{equation}\label{eq-notat}
\!\!\rev{\bz^{k}\!=\![z_1^{k},\ldots,z_m^{k}]^\T}\!\!, ~~
\hat{\bz}^{k}\!=\!\one_m(\hat{z}^{k})^\T\!\!,~~
\bx^*\!=\!\one_m(x^*)^\T\!\!,
\end{equation}
where $\hat{z}^k=\sum_{i=1}^m[\pi_k]_i z_i^k$
and $x^*$ is an NE point of the game. Then, the local update in compact form is as follows
\begin{equation}\label{eq-Compz}
\bz^{k+1} = W_k\bz^k - \bF_\a(W_k\bz^k)+\bbeta(\bz^k-\bz^{k-1}),
\end{equation}
where $\rev{\beta=(\beta_1,\ldots,\beta_m)^\T}$ 
and $\boldsymbol{\beta}=\Diag(\beta)$.

Let $\boldsymbol{\Pi}_{k}=\one_m \pi_{k}^\T$, for all $k \ge 0$. The weighted average $\hat{\bz}^{k}$ evolves according to the following relation
\begin{equation}\label{eq-Compzhat}
\hat{\bz}^{k+1} = \hat{\bz}^k - \boldsymbol{\Pi}_{k+1}\bF_\a(W_k\bz^k)+\boldsymbol{\Pi}_{k+1}\bbeta(\bz^k-\bz^{k-1}).
\end{equation}
We denote the following bounds:
\begin{equation}\label{eq-notat2}
\bar{\a}=\max\limits_{i\in[m]}\a_i,~ \munderbar{\a}=\min\limits_{i\in[m]}\a_i,~ \bar{\b}=\max\limits_{i\in[m]}\b_i.
\end{equation}

Let Assumption~\ref{assum:map_monotone}-\ref{asum-row-stochastic} hold. Consider Algorithm~\textbf{DNE-HB} and the notations in~\eqref{eq-notat}, \eqref{eq-notat2}. We have the following results:

\begin{proposition}\label{prop-wavgdist}
Let $c_k\!\!=\!\!\sqrt{1\!-\! \tfrac{\min(\pi_{k+1})w^2}{\max^2(\pi_k)\mathsf{D}(\G_k)\mathsf{K}(\G_k)}}$, we have
\begin{align*}
\|\bz^{k+1} &- \hat \bz^{k+1}\|_{\pi_{k+1}} \le  (1+\bL_\a) c_k \|\bz^k-\hat{\bz}^k\|_{\pi_k}\\
&+\bL_\a \|\hat \bz^k- \bx^* \|_{\pi_k}+ \bar{\b} \|\bz^k-\bz^{k-1} \|\!\! \text{for all}\!\! k\ge0.
\end{align*}
\end{proposition}
\begin{proof}
Using the update formulations in \eqref{eq-Compz} and \eqref{eq-Compzhat},
\begin{align}\label{eq-proofConsensusError1}
&\|\bz^{k+1} - \hat \bz^{k+1}\|_{\pi_{k+1}} \cr
= &\|W_k\bz^k \!\!-\! \hat{\bz}^k \!\!+\! (\mbi \!\!-\! \boldsymbol{\Pi}_{k+1}) \left(\bbeta(\bz^k \!\!-\!\bz^{k-1}\!) \!-\!\bF_\a(W_k\bz^k)\!\right)\|_{\pi_{k+1}}\cr
\le &\|W_k\bz^k - \hat{\bz}^k\|_{\pi_{k+1}}+\|(\mbi - \boldsymbol{\Pi}_{k+1}) \bF_\a(W_k\bz^k)\|_{\pi_{k+1}}\cr&+\|(\mbi - \boldsymbol{\Pi}_{k+1})\bbeta(\bz^k-\bz^{k-1})\|_{\pi_{k+1}}.
\end{align}

To evaluate the first term on the right-hand side (RHS) of \eqref{eq-proofConsensusError1}, we utilize Lemma~\ref{lemma-lemma6PushPull} with $W=W_k$, $z_i=z_i^k$, $u=\hat{z}^k$, and the stochastic vectors $\phi=\pi_{k+1}$ and $\pi=\pi_k$, to obtain
\begin{equation}\label{eq-piNormEstFromWeightedAvg}
\|W_k\bz^k - \hat{\bz}^k\|_{\pi_{k+1}}
     \le c_k \|\bz^k-\hat{\bz}^k\|_{\pi_k}\text{for all} k\ge0.
\end{equation} 

Consider the second term on the RHS of \eqref{eq-proofConsensusError1}. Let $\bg^k_i = \nabla_iJ_i([W_k\bz^k]_{i,:})$.
Since $\pi_{k+1}$ is a stochastic vector, we obtain
\begin{align*}
&\|(\mbi - \rev{\boldsymbol{\Pi}_{k+1}})\bF_\a(W_k\bz^k)\|^2_{\pi_{k+1}} \nonumber\\ 
\le &\sum_{i=1}^m [\pi_{k+1}]_i \Bigg(\!\!(1\!-\![\pi_{k+1}]_i)^2+[\pi_{k+1}]_i \!\!\sum_{j=1, j\neq i}^m \![\pi_{k+1}]_j\! \Bigg)\!\|\bg_i^k\|^2 \nonumber\\ 
\le & \sum_{i=1}^m [\pi_{k+1}]_i \left(1-[\pi_{k+1}]_i\right)\|\bg_i^k\|^2 \le  \sum_{i=1}^m [\pi_{k+1}]_i\|\bg_i^k\|^2.
\end{align*}
Notice that $\|\bF_\a(W_k\bz^k)\|^2_{\pi_{k+1}}=\sum_{i=1}^m [\pi_{k+1}]_i\|\bg_i^k\|^2$. Hence,
\begin{align}\label{eq-proofConsensusError2}
&\|(\mbi - \boldsymbol{\Pi}_{k+1}) \bF_\a(W_k\bz^k)\|_{\pi_{k+1}}\le   \|\bF_\a(W_k\bz^k) \|_{\pi_{k+1}} \nonumber\\
\!=& \rev{\|\bF_\a(W_k\bz^k)\!-\!\bF_\a(\bx^*) \|_{\pi_{k+1}}} \!\le \bL_\a\|W_k\bz^k\!-\!\bx^*\|_{\pi_{k+1}},
\end{align} 
where we use $\bF_\a(\bx^*)=\zero$ and Lemma~\ref{lemma-LipschitzbF}. 

Furthermore, we apply Lemma~\ref{lemma-lemma6PushPull} with $W=W_k$, $z_i=z_i^k$, $u=x^*$, and stochastic vectors $\phi=\pi_{k+1}$, $\pi=\pi_k$,
to obtain
\begin{equation}\label{eq-piNormEstFromNash}
\|W_k\bz^k \!-\bx^*\|_{\pi_{k+1}}^2 \!\le \|\bz^k \!- \bx^*\|_{\pi_k}^2 \!- (1-c_k^2) \|\bz^k \!-\hat{\bz}^k\|_{\pi_k}^2.
\end{equation}
By Lemma~\ref{lem-normlincomb}, with
$\g_i=[\pi_k]_i$, $u_i=z_i^k$ and $u=x^*$, and observing that 
$\|\hat z^k- x^* \|^2=\|\hat \bz^k- \bx^*\|^2_{\pi_k}$,
yields
\begin{equation}\label{eq-zaverandNash}
\|\bz^k-\bx^*\|^2_{\pi_k}
=\|\bz^k - \hat \bz^k\|^2_{\pi_k}
+\|\hat \bz^k- \bx^* \|^2_{\pi_k}.
\end{equation}
Combining the relations in \eqref{eq-piNormEstFromNash} and \eqref{eq-zaverandNash}, it follows that 
\begin{equation}\label{eq-piNormEstFromNash1}
\|W_k\bz^k -\bx^*\|_{\pi_{k+1}}
     \le \|\hat \bz^k- \bx^* \|_{\pi_k} +c_k \|\bz^k-\hat{\bz}^k\|_{\pi_k}.
\end{equation}
Combining the previous relation with \eqref{eq-proofConsensusError2}, we have
\begin{align}\label{eq-proofConsensusError2b}
\|(\mbi - &\boldsymbol{\Pi}_{k+1}) \bF_\a(W_k\bz^k)\|_{\pi_{k+1}}\nonumber\\
&\le \bL_\a \|\hat \bz^k- \bx^* \|_{\pi_k}+ \bL_\a c_k \|\bz^k-\hat{\bz}^k\|_{\pi_k}.
\end{align} 
For the last term on the RHS of \eqref{eq-proofConsensusError1}, let $\bu^k=\bz^k-\bz^{k-1}$, using Lemma~\ref{lem-normlincomb} with $\g_i=[\pi_{k+1}]_i$, vector $u_i$ as the $i$th row of the matrix $\bu$ (i.e., $u_i=\bu_{i:}^k$), and $u=0$, it follows that
\begin{align}\label{eq-proofConsensusError3}
&\|(\mbi \!- \boldsymbol{\Pi}_{k+1})\bbeta\bu^k\|^2_{\pi_{k+1}} \!\!= \!\!\sum_{i=1}^m [\pi_{k+1}]_i\b_i^2 \Bigg\|\bu_{i:}^k\!-\!\!\sum_{j=1}^m[\pi_{k+1}]_j\bu_{j:}^k\Bigg\|^2\nonumber\\ 
&\le \bar{\b}^2\sum_{i=1}^m [\pi_{k+1}]_i \|\bu^k_{i:}\|^2 = \bar{\b}^2|\bu^k\|^2_{\pi_{k+1}}\le\bar{\b}^2\|\bu^k\|^2.
\end{align}
The desired relation follows from \eqref{eq-proofConsensusError1}, \eqref{eq-piNormEstFromWeightedAvg}, \eqref{eq-proofConsensusError2b}, \eqref{eq-proofConsensusError3}.
\end{proof}

\begin{proposition}\label{prop-NEdist}
\!\rev{With $\alpha_i\!\in\!(0,2L_1^{-1})$ for all $i\!\in\!\![m]$,
we have}
\begin{align*}
\|\hat \bz^{k+1}& -\bx^*\|_{\pi_{k+1}} \le  \left(\bar{\a}\sqrt{2}L_2 +q_k(\a)\right)\|\hat \bz^k- \bx^* \|_{\pi_k} \nonumber\\
+&\sqrt{2}\bL_\a c_k \|\bz^k-\hat{\bz}^k\|_{\pi_k}+\bar{\b} \|\bz^k-\bz^{k-1} \| \!\!\text{for all}\!\! k\ge0,
\end{align*}
\rev{where $q_k(\a)\!=\!\max\limits_{i\in [m]}\{|1\!-[\pi_{k+1}]_i\a_i\mu_i|,|1\!-[\pi_{k+1}]_i\a_i L_i|\}\!<\!1$, $L_1=\max\limits_{i\in[m]}{L_i}$ and $L_{2}=\max\limits_{i\in[m]}{L_{-i}}$.}
\end{proposition}
\begin{proof}
Using the compact form for $\hat{\bz}^{k+1}$ in \eqref{eq-Compzhat},
\begin{align}\label{eq-zhatxstar}
&\|\hat \bz^{k+1} -\bx^*\|_{\pi_{k+1}} \nonumber\\
\le& \|\hat{\bz}^k \!\!-\! \bx^*\!\!-\!\boldsymbol{\Pi}_{k+1}\bF_\a(W_k\bz^k\!)\|_{\pi_{k+1}}\!\!+\!\|\boldsymbol{\Pi}_{k+1}\bbeta(\bz^k\!\!-\!\bz^{k-1}\!)\|_{\pi_{k+1}}\nonumber\\
\le& \|\hat{\bz}^k \!\!-\! \bx^* \!\!-\!\boldsymbol{\Pi}_{k+1}\bF_0(\hat{\bz}^k,\bx^*\!)\|_{\pi_{k+1}}\!\!+\!\|\boldsymbol{\Pi}_{k+1}\bbeta(\bz^k\!\!-\!\bz^{k-1}\!)\|_{\pi_{k+1}}\nonumber\\
&+\|\boldsymbol{\Pi}_{k+1}\bF_0(\hat{\bz}^k,\bx^*)-\boldsymbol{\Pi}_{k+1}\bF_\a(W_k\bz^k)\|_{\pi_{k+1}},
\end{align}
where the matrix $\bF_0(\hat{\bz}^k\!,\bx^*\!)$ has $i$th row, $[\bF_0(\hat{\bz}^k\!,\bx^*\!)]_{i:}$, equals
\begin{align}\label{eq-def-bf0}
(\zero^\T_{n_1}\!,\ldots, \zero^\T_{n_{i-1}}\!, \a_i(\nabla_i J_i(\hat z_i^k,x_{-i}^*))^\T\!, \zero^\T_{n_{i+1}}\!,\ldots,\zero^\T_{n_m}\! ).\!
\end{align}


For the first term on the RHS of \eqref{eq-zhatxstar}, since the matrices are consensual and $\pi_{k+1}$ is a stochastic vector, we have
\begin{align*}
\|\hat{\bz}^k -& \bx^*-\boldsymbol{\Pi}_{k+1}\bF_0(\hat{\bz}^k,\bx^*)\|^2_{\pi_{k+1}}\nonumber\\
&= \sum_{i=1}^m\|\hat z_i^{k}-x_i^*-[\pi_{k+1}]_i\a_i\nabla_i J_i(\hat z_i^k,x_{-i}^*)\| ^2.
\end{align*}
Applying Lemma~\ref{lem-contraction}, 
for all $\a_i$ such that $0\!\!<\!\![\pi_{k+1}]_i\a_i\!\!<\!\!2L_i^{-1}$:
\begin{align*}
&\|\hat z_i^{k}-x_i^*-[\pi_{k+1}]_i\a\nabla_i J_i(\hat z_i^k,x_{-i}^*)\| \le q_{i,k}(\a) \|\hat z_i^{k}-x_i^*\|,
\end{align*}
$q_{i,k}(\a)\!=\!\max\{|1\!-\![\pi_{k+1}]_i\a_i\mu_i|,|1\!-\![\pi_{k+1}]_i\a_i L_i|\}$. Hence,
\begin{align}\label{eq-zhatxstar1}
\|\hat{\bz}^k &- \bx^*-\boldsymbol{\Pi}_{k+1}\bF_0(\hat{\bz}^k,\bx^*)\|^2_{\pi_{k+1}} \le q^2_k(\a)\sum_{i=1}^m\|\hat z_i^{k}-x_i^*\|^2\nonumber\\
&=q^2_k(\a)\|\hat{z}^{k} -x^*\|^2 = q^2_k(\a)\|\hat \bz^{k} -\bx^*\|^2_{\pi_{k}}.
\end{align}

Regarding the second term on the RHS of \eqref{eq-zhatxstar}, the matrix $\boldsymbol{\Pi}_{k+1}\bbeta(\bz^k-\bz^{k-1})$ is consensual. For stochastic vectors $\pi_{k+1}$, using the notation $\bu^k=\bz^k-\bz^{k-1}$, we have
\begin{align}\label{eq-zhatxstar2}
\|\boldsymbol{\Pi}_{k+1}\bbeta\bu^k\|^2_{\pi_{k+1}}\!\!&=\!\bigg\|\sum_{i=1}^m[\pi_{k+1}]_i\b_i\bu^k_{i:}\bigg\|^2\!\!\!\le \bar{\b}\sum_{i=1}^m[\pi_{k+1}]_i\|\bu^k_{i:}\|^2 \nonumber\\
&\le \bar{\b}\|\bu^k\|^2=\bar{\b}\|\bz^k-\bz^{k-1}\|^2,
\end{align}
where we use Lemma~\ref{lem-normlincomb} with $\g_i=[\pi_{k+1}]_i$, $u_i=\bu_{i:}^k$, and $u\!=\!0$.
For the last term in \eqref{eq-zhatxstar}, from Assumption~\ref{assum:lip} we obtain
\begin{align*}
&\|\boldsymbol{\Pi}_{k+1}\bF_0(\hat{\bz}^k,\bx^*)-\boldsymbol{\Pi}_{k+1}\bF_\a(W_k\bz^k)\|^2_{\pi_{k+1}}\nonumber\\
=&\bar{\a}^2\sum_{i=1}^m[\pi_{k+1}]^2_i\|\nabla_i J_i(\hat z_i^k,x_{-i}^*)-\nabla_i J_i([W_k\bz^k]_i)\|^2 \nonumber\\
\le&\bar{\a}^2\sum_{i=1}^m[\pi_{k+1}]^2_i( 2\|\nabla_i J_i(\hat z_i^k\!,x_{-i}^*)-\nabla_i J_i([W_k\bz^k]_{ii},x_{-i}^*)\|^2 \nonumber\\
&\qquad \qquad +2\|\nabla_i J_i([W_k\bz^k]_{ii},x_{-i}^*)-\nabla_i J_i([W_k\bz^k]_i)\|^2)\nonumber\\
\le & \bar{\a}^2 (2L_1^2\|\hat \bz^k-W_k\bz^k\|^2_{\pi_{k+1}}+2 L_{2}^2\|\bx^*-W_k\bz^k\|^2_{\pi_{k+1}}).
\end{align*}

Combining the relations in \eqref{eq-piNormEstFromWeightedAvg} and \eqref{eq-piNormEstFromNash1} with the preceding relation and using $\sqrt{a\!+\!b}\!\le \!\sqrt{a}\!+\!\sqrt{b}, \forall a,b\!\ge\!0$, yields
\begin{align}\label{eq-zhatxstar3}
\!\!\!\|\boldsymbol{\Pi}_{k+1}&\bF_0(\hat{\bz}^k,\bx^*)-\boldsymbol{\Pi}_{k+1}\bF_\a(W_k\bz^k)\|\nonumber\\
&\!\!\!\!\!\le \bar{\a} \sqrt{2} L_2\|\hat \bz^k- \bx^* \|_{\pi_k} +\sqrt{2}\bL_\a c_k\|\bz^k-\hat{\bz}^k\|_{\pi_k},\!
\end{align}
where we use the relation $ \bar{\a} L_1\sqrt{2}+\bar{\a}L_2\sqrt{2}\le \sqrt{2}\bL_\a$, by the Cauchy–Schwarz inequality, to obtain the last term.

The desired relation follows from \eqref{eq-zhatxstar} and \eqref{eq-zhatxstar1}-\eqref{eq-zhatxstar3}.
\end{proof}

\begin{proposition}\label{prop-statediff}
Let $\varphi_{k}=\frac{1}{\sqrt{\min\pi_{k}}}$. We then have
\begin{align*}
\|&\bz^{k+1}-\bz^{k} \| \le \varphi_{k+1}(1+\bL_\a) c_k \|\bz^k-\hat{\bz}^k\|_{\pi_k}  \nonumber\\
&+\varphi_{k+1}\bL_\a\|\hat \bz^k- \bx^* \|_{\pi_k}+\bar{\b} \|\bz^k-\bz^{k-1} \| \text{for all} k\ge0.
\end{align*}
\end{proposition}
\begin{proof}
Using the update formulation in \eqref{eq-Compz}, we obtain
\begin{align}\label{eq-statediffrel}
&\!\!\|\bz^{k+1}-\bz^{k}\| \le \|W_k\bz^k \!- \bF_\a(W_k\bz^k)+\bbeta(\bz^k\!-\bz^{k-1})\!-\bz^{k} \|\nonumber\\
&\!\!\le \|W_k\bz^k -\bz^{k} \|+ \|\bF_\a(W_k\bz^k)\|+\|\bbeta(\bz^k\!-\bz^{k-1})\|.
\end{align}

Using the relations in \eqref{eq-NormIneq} and \eqref{eq-piNormEstFromWeightedAvg}, we have:
\begin{align}\label{eq-statediffrel1}
\|W_k\bz^k - \hat{\bz}^k\| \le \varphi_{k+1} c_k \|\bz^k-\hat{\bz}^k\|_{\pi_k}.
\end{align} 
From relations \eqref{eq-NormIneq} and since $\bF_\a(\bx^*)=\zero$, we see that
\begin{align}\label{eq-statediffrel2}
&\|\bF_\a(W_k\bz^k)\| \le \varphi_{k+1}\|\bF_\a(W_k\bz^k)-\bF_\a(\bx^*) \|_{\pi_{k+1}}\nonumber\\
\le & ~\bL_\a\varphi_{k+1}\left(\|\hat \bz^k- \bx^* \|_{\pi_k} + c_k \|\bz^k-\hat{\bz}^k\|_{\pi_k}\right),
\end{align} 
where we use Lemma~\ref{lemma-LipschitzbF} and relation \eqref{eq-piNormEstFromNash1}.
For the last term on the RHS of \eqref{eq-statediffrel}, we have
\begin{align}\label{eq-statediffrel3}
\|\boldsymbol{\beta}(\bz^k-\bz^{k-1})\|\le \bar{\beta} \|\bz^k-\bz^{k-1} \|.
\end{align} 
The desired relation follows from \eqref{eq-statediffrel}-\eqref{eq-statediffrel3}.
\end{proof}


Define $V_k=\left(\|\bz^{k} - \hat \bz^{k}\|_{\pi_{k}},\|\hat \bz^{k} -\bx^*\|_{\pi_{k}},\|\bz^{k}\!-\!\bz^{k-1}\|\right)^{\!\T}$, we  have the following composite relation:

\begin{proposition}\label{prop-comprel}
\!With $\alpha_i\!\in\!(0,2L_1^{-1})$ for all $i\!\in\!\![m]$, 
we have 
\[V_{k+1}\le M_k(\a,\b)V_k\qquad \hbox{for all $k\ge0$},\]
with $\displaystyle M_k(\a,\b)\!=\!\!\begin{bmatrix}
(1+\bL_\a) c_k & \bL_\a & \bar{\b}\\
\sqrt{2}\bL_\a c_k  & \!\!\!\bar{\a}\sqrt{2}L_2 +q_k(\a) & \bar{\b}\\
\varphi_{k+1}(1+\bL_\a) c_k & \varphi_{k+1}\bL_\a & \bar{\b}
\end{bmatrix}\!\!,$\\
where $\bL_\a$ defined as in Lemma~\ref{lemma-LipschitzbF}, $q_k(\a)$ defined as in Proposition~\ref{prop-NEdist}, 
and $\varphi_{k+1}$ defined as in Proposition~\ref{prop-statediff}. 
\end{proposition}
\begin{proof}
The relation follows from Propositions~\ref{prop-wavgdist}-\ref{prop-statediff}.
\end{proof}

In view of Proposition~\ref{prop-comprel}, to prove that $V_k\to0$ at a geometric rate,  it suffices to show that $M_k(\a,\b)\le M(\a,\b)$,
for some matrix $M(\a,\b)$. Then, we select appropriate step-size and the momentum parameter such that the spectral radius $\rho_{M}$ of $M(\a,\b)$ is less than $1$, as follows.
We have \rev{$\bL_\a \le \bar{\a}L$, where $L= \sqrt{L_1^2+L_2^2}$ with $L_1$ and $L_2$ as in Proposition~\ref{prop-NEdist}.} Additionally, for $\bar{\a}\in(0,2(L_1+\mu)^{-1})$, we obtain $q_k(\a)=1-\munderbar{\a} \min(\pi_k)\mu <1$. We \rev{let $\sigma>0$ be such that $\sigma \le \min_{k\ge0}{\min(\pi_k)}$ (see Lemma~\ref{lemma-TimeVaryPiVectors}).}
Thus, we have
\begin{align}\label{eq-maxconst}
	\max_{k\ge0}q_k(\a) \le 1-\munderbar{\a}\sigma \mu,~\max_{k\ge0}c_k\le c,~ \max_{k\ge0} \varphi_k \le \varphi.
\end{align}
Using the bounds in~\eqref{eq-notat2} and \eqref{eq-maxconst}, for $\a\in(0,2(L_1+\mu)^{-1})$, we have $M_k(\a,\b)\le M(\a,\b)$ for all $k \ge 0$, with 
\begin{align}\label{eq-gmatrixm}
M(\a,\b)\!=\!\begin{bmatrix}
(1+\bar{\a}L) c & \bar{\a}L & \bar{\b}\\
\sqrt{2}\bar{\a}L c  & \!\!1-(\munderbar{\a}\sigma\mu-\bar{\a}\sqrt{2}L_2) & \bar{\b}\\
\varphi(1+\bar{\a}L) c & \varphi\bar{\a}L & \bar{\b}
\end{bmatrix}\!\!.\!
\end{align}

We now provide the main convergence result.
\begin{theorem} \label{theo:convTheo}
Let Assumptions~\ref{assum:map_monotone}-\ref{asum-row-stochastic} hold, and assume that $\sigma\mu>\sqrt{2}L_2$. 
For all $i\in[m]$, let the step-size $\a_i>0$ and the acceleration parameter $\b_i\ge 0$ be such that 
\begin{align}\label{eq:alpha-range}
	\!\!\bar{\a} \le \min \!\Big\{\!\tfrac{2}{L_1+\mu},\tfrac{1-c}{Lc},\tfrac{\eta_1}{\eta_2}\!\Big\}\!,\munderbar{\a}\!>\!\tfrac{\bar{\a}\sqrt{2}L_2}{\sigma\mu},\bar{\b}\!<\! \tfrac{\eta(1-c)}{(2-c)(\eta+L)},
\end{align}
where $\eta\bar{\a}=\munderbar{\a}\sigma\mu-\bar{\a}\sqrt{2}L_2$, $\eta_1\!=\eta(1\!-\!c)-\bar{\b}(2\!-\!c)(\eta\!+\!L)\!>\!0$, and $\eta_2\!=\!\sqrt{2}Lc(1\!+\!\bar{\b}(\varphi\!-\!c))\!+\!\bar{\b}c\eta(\varphi\!-\!1)\!+\!c\eta\!>\!0$.
Then, $\rho_{M}<1$, thus, $\lim_{k\to\infty}\|\bz^k-\bx^*\|=0$ and $\lim_{k\to\infty}\|x^k-x^*\|=0$ with a linear convergence rate of the order of $\mathcal{O}\Big(\rho_M^k\Big)$.
\end{theorem}

\begin{proof}
By Lemma 8 in \cite{pshi21}, we obtain $\rho_{M}\!<\!1$ if $\det(\mbi\!-\!M(\a,\b))\!>\!0$ and the diagonal entries of $M(\a,\b)$ are less than $1$. Solving the resulting inequalities yields \eqref{eq:alpha-range}. 
\end{proof}
 


\begin{remark}\label{remark-ConditionStrongF}
The assumption $\sigma\mu>\sqrt{2}L_2$ is equivalent to the condition that $\b_\a>0$ in \cite{nguyen2022distributed}, establishing when $\bF_\a(\cdot)$ is strongly monotone. \rev{This assumption is akin to those  in Proposition 7 of \cite{Scutari2014}, Lemma 2 of \cite{Tatiana2018}, and Assumption 5 in \cite{Tatarenko2021}. Essentially, it entails  that the strong monotonicity of each agent's local objective is sufficiently strong compared to its dependence on the actions of other agents or the coupling of the agents' optimization problems, to guarantee the strong monotonicity of the game mapping. In Section VI of~\cite{Scutari2014}, this assumption has a compelling physical interpretation in the context of power control  in cognitive radio networks.} 
\end{remark}

\section{Numerical results} \label{sec:simulation}

We evaluate the performance of the proposed approach for a  Nash-Cournot game, as described in \cite{nguyen2022distributed}. Consider $m=20$ firms competing in $N=7$ markets, denoted as $M_1,\ldots,M_N$. Each firm $i\in [m]$ competes in $n_i\le N$ markets by determining the quantity of the homogeneous commodity $x_i \in \Omega_i =\R^{n_i}$ to be produced and delivered, as illustrated in Figure 1 of \cite{nguyen2022distributed}.
Firm $i$ has a local matrix $B_i\in\R^{N\times n_i}$, with $[B_i]_{hj}= 1$ if agent i delivers $[x_i]_j$ to $M_h$, $h \in [N]$, and $0$ otherwise. Let $n=\sum_{i=1}^m n_i=32$, $x=[x_i]_{i\in [m]}\in\R^n$, and $B=[B_1,\cdots,B_m]\in\R^{N\times n}$. Then, given an action profile $x$ of all the firms, the vector of the total product supplied to the markets can be expressed as $Bx=\sum_{i=1}^mB_ix_i\in \R^N$.
The commodity's price in $M_h$ is
$p_h(x)=\Bar{P}_h-\chi_h[Bx]_h$, $\forall h$,
where $\Bar{P}_h>0$ and $\chi_h>0$. Let $\Bar{P}=[\Bar{P}_h]_{h=\overline{1,N}}\in\R^N$ and $\Xi = \diag([\chi_h]_{h=\overline{1,N}})\in\R^{N\times N}$. Then, the price vector function $P=[p_h]_{h=\overline{1,N}}$ has the form:
$P=\Bar{P}-\Xi Bx$, and $P^\T B_ix_i$ is the payoff of firm $i$ obtained by selling $x_i$  to the markets that it connects with.
Firm $i$’s production cost is 
$c_i(x_i)=x_i^\T Q_ix_i+q_i^\T x_i,$
with $Q_i\in \R^{n_i\times n_i}$ symmetric and $Q_i\succ 0$, and $q_i\in\R^{n_i}$.
The objective function of firm $i$ is
  $  J_i(x_i,x_{-i})=c_i(x_i)-(\Bar{P}-\Xi Bx)^\T B_ix_i$.
In our simulations, we use directed time-varying graphs with self-loops and establish a directed cycle linking all agents at each iteration. We define the row-stochastic weight matrix $W_k$ as in \cite{nguyen2022distributed}, generate the diagonal matrix $Q_i$ with entries uniformly distributed in $[5,8]$, draw $q_i$ uniformly from the interval $[1,2]$ and select $\Bar{P}_h$ randomly from $[10,20]$. We choose $\chi_h$ such that $\sigma\mu>\sqrt{2}L_2$, for example, $0.01\le \chi_h \le 0.02$ yields $\mu$ around $11$ and $L_2$ around $0.03$, indicating that the strong monotonicity of the local objective is significantly strong in comparison to its dependence on other agents' actions.

\begin{figure}[t!]
\vspace{0.2cm}
	\centering
	\hspace*{-1.7em}
	\includegraphics[width=0.23\textwidth]{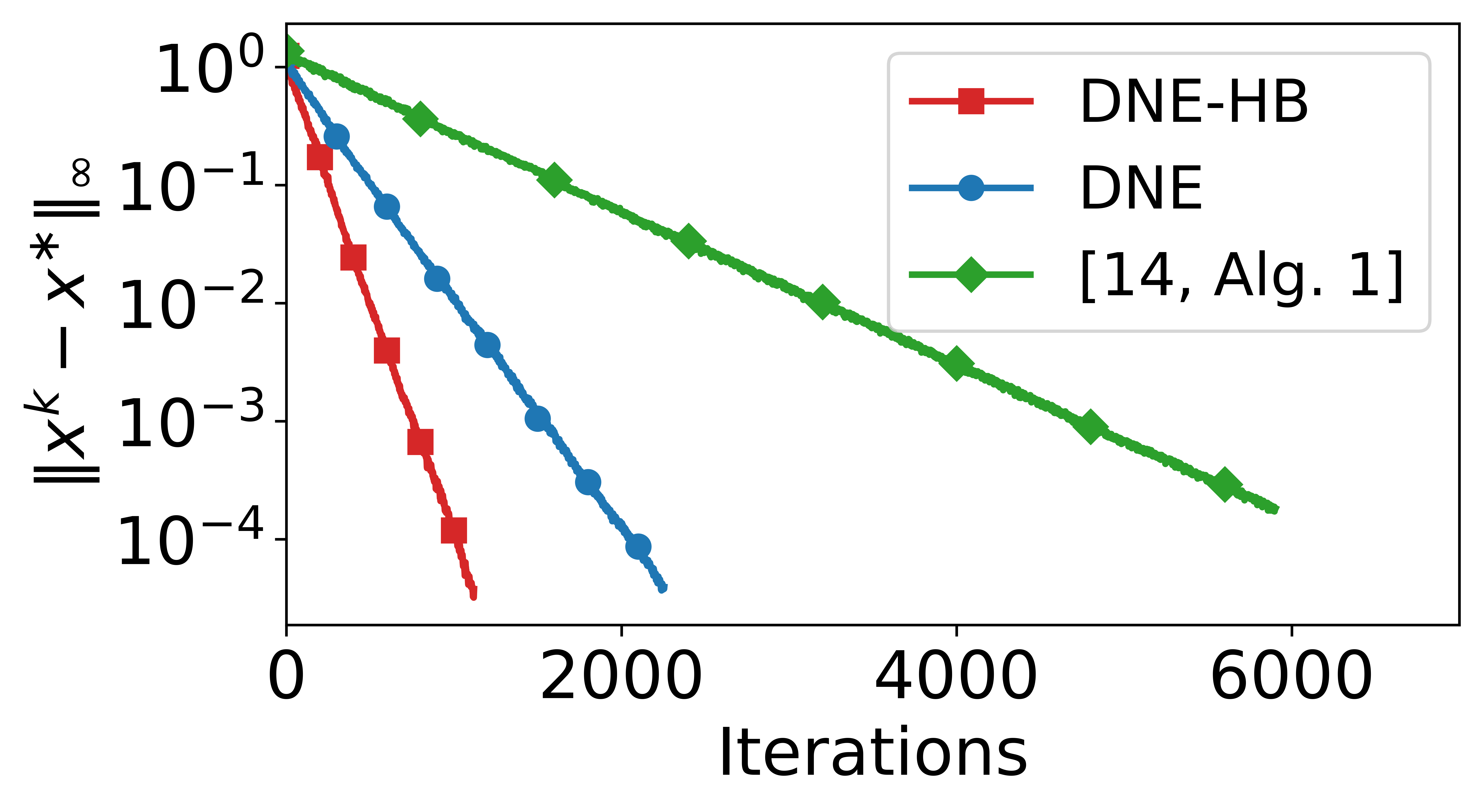}
	\includegraphics[width=0.23\textwidth]{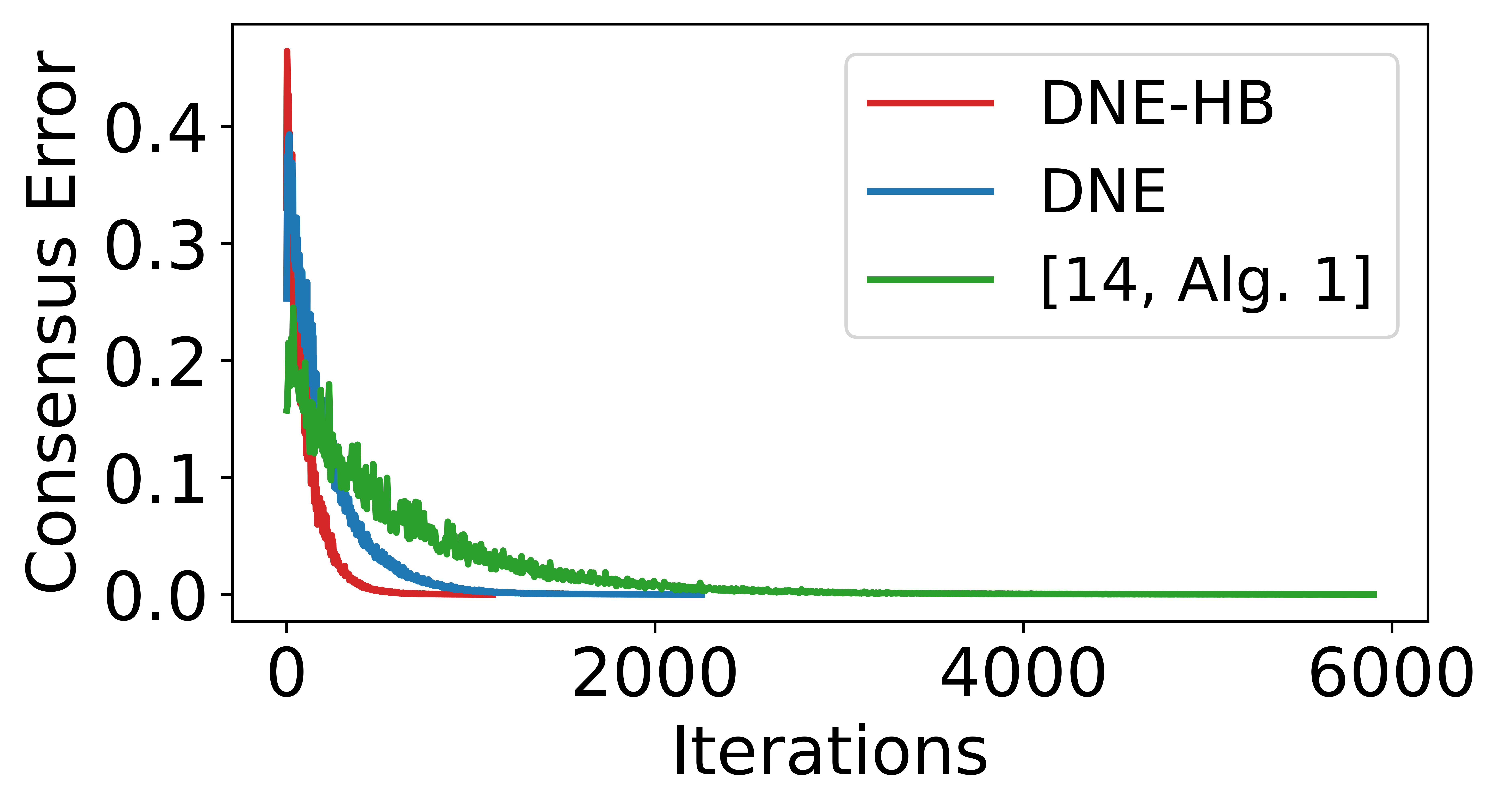}
	\hspace*{-1.7em}
    \vspace{-0.2cm}
	\caption{Convergence behavior for one game instance.}
	\label{fig:errors}
	\vspace{-0.2cm}
\end{figure}
\begin{figure}[t!]
	\centering
	\includegraphics[width=0.23\textwidth]{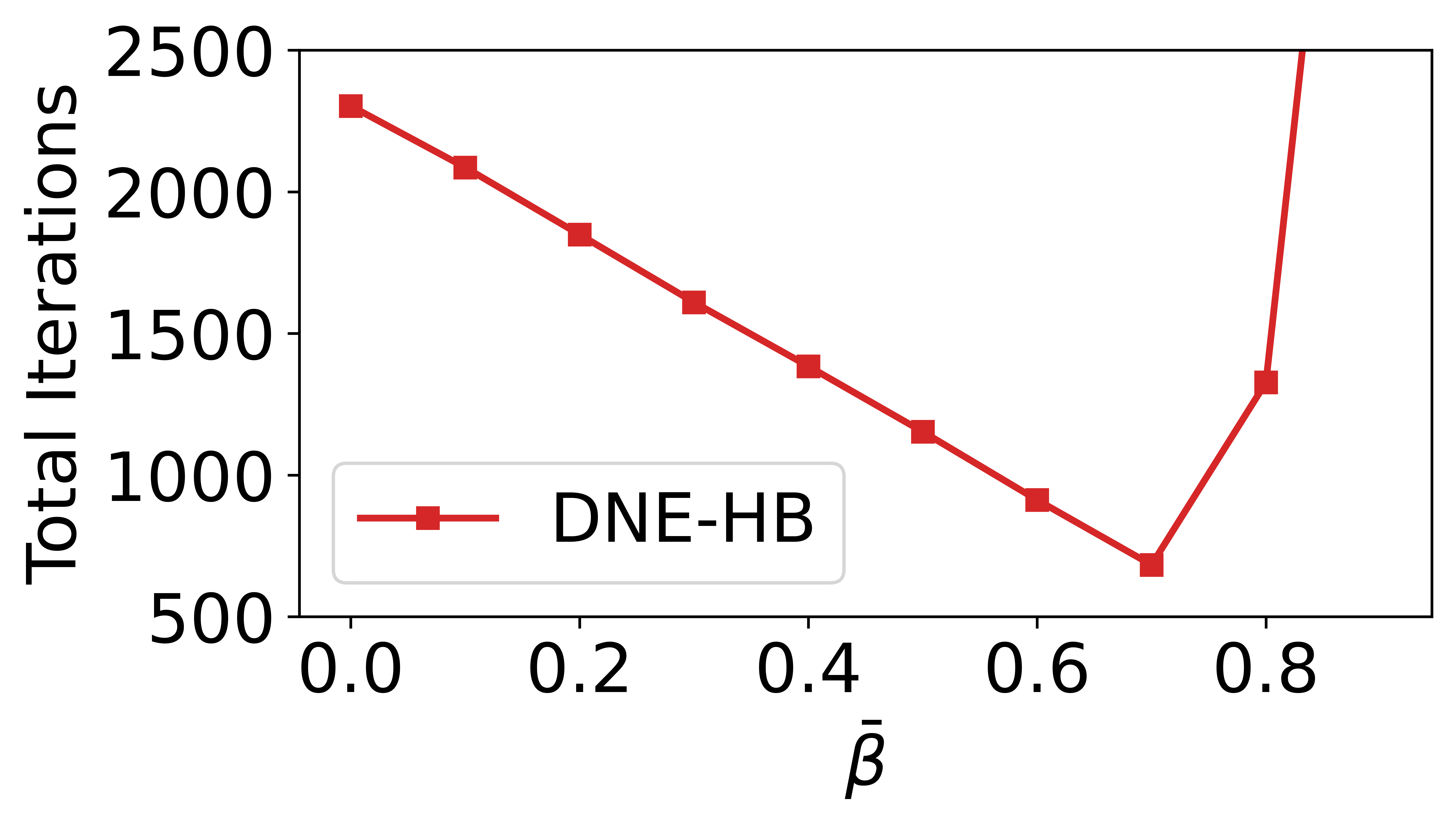}\vspace{-0.2cm}
	\caption{Effects of varying the momentum parameter.}\label{fig:effects}
 \vspace{-0.2cm}
\end{figure}


To demonstrate the accelerated convergence of \textbf{DNE-HB}, we compare it with existing agorithms including \textbf{DNE} (i.e. Algo.~1 of~\cite{nguyen2022distributed}) and Algo.~1 of~\cite{Bianchi2020NashES}. We conduct $1000$ simulations with $\alpha_i=0.01$, $\beta_i=0.5$ and terminate the algorithms if the consensus error $\max_{i\in[m],j\in[m],i\neq j} \|[\bz_{k}]_{i:}-[\bz_k]_{j:}\|_\infty$ is less than $10^{-5}$ or the iteration limit of $10^5$ is reached. Figure~\ref{fig:errors} shows the convergence for a game instance. Figure~\ref{fig:effects} illustrates the effect of momentum parameter on convergence rate. These results suggest that the algorithm converges faster with larger momentum parameter values satisfying the range in~\eqref{eq:alpha-range}. Table~\ref{table:Simulations} further compares the average performance of the algorithms over $1000$ simulations. Overall, the results demonstrate that the proposed algorithm  with the heavy-ball acceleration significantly improves the convergence rate.
\begin{table}[t!]\centering
\begin{tabular}{|c|c|c|c|c|}
\hline
\multirow{2}{*}{Stepsize} & \multicolumn{2}{c|}{Avg. \# Iterations}           & \multicolumn{2}{c|}{Avg. Running Time (s)}    \\ \cline{2-5} 
&\textbf{DNE}    & \textbf{DNE-HB}  &  \textbf{DNE}   & 	\textbf{DNE-HB}  \\ \hline
$\alpha_i=0.005$ & $3667.52$ & $1826.44$ & $1.0703$ & $0.5984$      \\ \hline
$\alpha_i=0.01$ & $2032.25$ & $1006.56$ & $0.6702$ & $0.3502$    \\ \hline
\end{tabular}
\caption{Average performance over $1000$ simulations.}
\label{table:Simulations} 
\vspace{-0.4cm}
\end{table}

\section{Conclusions and Future Work}
\label{sec:conc}
\rev{This paper has proposed an accelerated distributed algorithm that incorporates heavy-ball acceleration to improve the performance of the gradient-based distributed NE-seeking algorithm for games over time-varying directed networks. A geometric convergence rate of the algorithm is shown with explicit bounds for the non-identical step-sizes and momentum parameters based on the properties of the cost functions and network structure. Our numerical results illustrate the effectiveness  of the proposed method.
In future, we will study its convergence for games with constrained action sets.}

\bibliographystyle{IEEEtran}
\bibliography{references}

\begin{thebibliography}{10}
\providecommand{\url}[1]{#1}
\csname url@samestyle\endcsname
\providecommand{\newblock}{\relax}
\providecommand{\bibinfo}[2]{#2}
\providecommand{\BIBentrySTDinterwordspacing}{\spaceskip=0pt\relax}
\providecommand{\BIBentryALTinterwordstretchfactor}{4}
\providecommand{\BIBentryALTinterwordspacing}{\spaceskip=\fontdimen2\font plus
\BIBentryALTinterwordstretchfactor\fontdimen3\font minus
  \fontdimen4\font\relax}
\providecommand{\BIBforeignlanguage}[2]{{%
\expandafter\ifx\csname l@#1\endcsname\relax
\typeout{** WARNING: IEEEtran.bst: No hyphenation pattern has been}%
\typeout{** loaded for the language `#1'. Using the pattern for}%
\typeout{** the default language instead.}%
\else
\language=\csname l@#1\endcsname
\fi
#2}}
\providecommand{\BIBdecl}{\relax}
\BIBdecl

\bibitem{BasharSG}
W.~Saad, Z.~Han, H.~V. Poor, and T.~Basar, ``Game-{T}heoretic {M}ethods for the
  {S}mart {G}rid: {A}n {O}verview of {M}icrogrid {S}ystems, {D}emand-{S}ide
  {M}anagement, and {S}mart {G}rid {C}ommunications,'' \emph{IEEE Signal
  Process. Mag.}, vol.~29, no.~5, pp. 86--105, 2012.

\bibitem{wiopt23}
D.~T.~A. Nguyen, J.~Cheng, D.~T. Nguyen, and A.~Nedi\'c, ``Crowd{C}ache: {A}
  {D}ecentralized {G}ame--{T}heoretic {F}ramework for {M}obile {E}dge {C}ontent
  {S}haring,'' \emph{arXiv preprint arXiv:2304.13246}, 2023.

\bibitem{Yi2019}
P.~Yi and L.~Pavel, ``An {O}perator {S}plitting {A}pproach for {D}istributed
  {G}eneralized {N}ash {E}quilibria {C}omputation,'' \emph{Automatica}, vol.
  102, pp. 111--121, 2019.

\bibitem{Belgioioso2018}
G.~Belgioioso and S.~Grammatico, ``Projected-{G}radient {A}lgorithms for
  {G}eneralized {E}quilibrium {S}eeking in {A}ggregative {G}ames are
  {P}reconditioned {F}orward-{B}ackward {M}ethods,'' in \emph{2018 European
  Control Conference (ECC)}, 2018, pp. 2188--2193.

\bibitem{Gadjov2019}
D.~Gadjov and L.~Pavel, ``A {P}assivity-{B}ased {A}pproach to {N}ash
  {E}quilibrium {S}eeking {O}ver {N}etworks,'' \emph{IEEE Trans. Autom.
  Control}, vol.~64, no.~3, pp. 1077--1092, 2019.

\bibitem{Koshal2016}
J.~Koshal, A.~Nedi\'c, and U.~V. Shanbhag, ``Distributed {A}lgorithms for
  {A}ggregative {G}ames on {G}raphs,'' \emph{Operations Research}, vol.~64,
  no.~3, pp. 680--704, 2016.

\bibitem{Tatiana2020}
T.~Tatarenko and A.~Nedi\'c, ``Geometric {C}onvergence of {D}istributed
  {G}radient {P}lay in {G}ames with {U}nconstrained {A}ction {S}ets,''
  \emph{IFAC-PapersOnLine}, vol.~53, pp. 3367--3372, 01 2020.

\bibitem{SALEHISADAGHIANI201927}
F.~Salehisadaghiani, W.~Shi, and L.~Pavel, ``Distributed {N}ash {E}quilibrium
  {S}eeking {U}nder {P}artial-{D}ecision {I}nformation via the {A}lternating
  {D}irection {M}ethod of {M}ultipliers,'' \emph{Automatica}, vol. 103, pp.
  27--35, 2019.

\bibitem{Tatiana2018}
T.~Tatarenko, W.~Shi, and A.~Nedi\'c, ``Accelerated {G}radient {P}lay
  {A}lgorithm for {D}istributed {N}ash {E}quilibrium {S}eeking,'' in \emph{2018
  IEEE Conf. Decis. Control (CDC)}, 2018, pp. 3561--3566.

\bibitem{Tatarenko2021}
------, ``Geometric {C}onvergence of {G}radient {P}lay {A}lgorithms for
  {D}istributed {N}ash {E}quilibrium {S}eeking,'' \emph{IEEE Trans. Autom.
  Control}, vol.~66, no.~11, pp. 5342--5353, 2021.

\bibitem{Grammatico2021}
G.~Belgioioso, A.~Nedi\'c, and S.~Grammatico, ``Distributed {G}eneralized
  {N}ash {E}quilibrium {S}eeking in {A}ggregative {G}ames on {T}ime-{V}arying
  {N}etworks,'' \emph{IEEE Trans. Autom. Control}, vol.~66, no.~5, pp.
  2061--2075, 2021.

\bibitem{Farzad2019}
F.~Salehisadaghiani and L.~Pavel, ``Nash {E}quilibrium {S}eeking with
  {N}on-doubly {S}tochastic {C}ommunication {W}eight {M}atrix,'' \emph{EAI
  Endorsed Trans. Collaborative Computing}, vol.~4, no.~13, pp. 3--15, 2019.

\bibitem{Bianchi_2021}
M.~Bianchi and S.~Grammatico, ``Fully {D}istributed {N}ash {E}quilibrium
  {S}eeking {O}ver {T}ime-{V}arying {C}ommunication {N}etworks {W}ith {L}inear
  {C}onvergence {R}ate,'' \emph{IEEE Control Syst. Lett.}, vol.~5, no.~2, pp.
  499--504, 2021.

\bibitem{Bianchi2020NashES}
------, ``Nash {E}quilibrium {S}eeking under {P}artial-{D}ecision {I}nformation
  {O}ver {D}irected {C}ommunication {N}etworks,'' in \emph{2020 59th IEEE Conf.
  Decis. Control (CDC)}, 2020, pp. 3555--3560.

\bibitem{nguyen2022distributed}
D.~T.~A. Nguyen, D.~T. Nguyen, and A.~Nedi{\'c}, ``{D}istributed {N}ash
  {E}quilibrium {S}eeking over {T}ime-{V}arying {D}irected {C}ommunication
  {N}etworks,'' \emph{arXiv preprint arXiv:2201.02323}, 2022.

\bibitem{GadjovTCSN2023}
D.~Gadjov and L.~Pavel, ``An {A}lgorithm for {R}esilient {N}ash {E}quilibrium
  {S}eeking in the {P}artial {I}nformation {S}etting,'' \emph{IEEE Transactions
  on Control of Network Systems}, pp. 1--10, 2023.

\bibitem{Polyak}
B.~Polyak, \emph{Introduction to Optimization}.\hskip 1em plus 0.5em minus
  0.4em\relax New York : Optimization Software, Inc., 1987.

\bibitem{Nguyen2023AccAB}
D.~T.~A. Nguyen, D.~T. Nguyen, and A.~Nedi{\'c}, ``Accelerated
  {AB}/{P}ush-{P}ull {M}ethods for {D}istributed {O}ptimization over
  {T}ime-{V}arying {D}irected {N}etworks,'' \emph{arXiv preprint
  arXiv:2302.01214}, 2023.

\bibitem{Song2020}
C.~Song, C.~Wu, Z.~Lv, F.~Zhang, J.~Li, and S.~Yang, ``Distributed
  {H}eavy-{B}all {N}ash {E}quilibrium {S}eeking {A}lgorithm in {A}ggregative
  {G}ames,'' in \emph{39th Chinese Control Conference}, 2020, pp. 5019--5024.

\bibitem{Fang2022}
X.~Fang, G.~Wen, J.~Zhou, J.~Lu, and G.~Chen, ``Distributed {N}ash
  {E}quilibrium {S}eeking for {A}ggregative {G}ames {W}ith {D}irected
  {C}ommunication {G}raphs,'' \emph{IEEE Trans. Circuits Syst. I: Regul. Pap.},
  vol.~69, no.~8, pp. 3339--3352, 2022.

\bibitem{BelgioiosoTAC2023}
G.~Belgioioso and S.~Grammatico, ``Semi-{D}ecentralized {G}eneralized {N}ash
  {E}quilibrium {S}eeking in {M}onotone {A}ggregative {G}ames,'' \emph{IEEE
  Trans. Autom. Control}, vol.~68, no.~1, pp. 140--155, 2023.

\bibitem{BIANCHI2022110080}
M.~Bianchi, G.~Belgioioso, and S.~Grammatico, ``Fast {G}eneralized {N}ash
  {E}quilibrium {S}eeking under partial-{D}ecision {I}nformation,''
  \emph{Automatica}, vol. 136, p. 110080, 2022.

\bibitem{Angelia2022AB}
A.~Nedi\'c, D.~T.~A. Nguyen, and D.~T. Nguyen, ``{AB}/{P}ush-{P}ull {M}ethod
  for {D}istributed {O}ptimization in {T}ime-{V}arying {D}irected {N}etworks,''
  \emph{arXiv preprint arXiv:2209.06974}, 2022.

\bibitem{FacchineiPang}
F.~Facchinei and J.-S. Pang, \emph{Finite-{D}imensional {V}ariational
  {I}nequalities and {C}omplementarity {P}roblems}.\hskip 1em plus 0.5em minus
  0.4em\relax Springer Series in Operations Research and Financial Engineering,
  2003.

\bibitem{pshi21}
S.~Pu, W.~Shi, J.~Xu, and A.~Nedi{\'c}, ``{P}ush–{P}ull {G}radient {M}ethods
  for {D}istributed {O}ptimization in {N}etworks,'' \emph{IEEE Trans. Autom.
  Control}, vol.~66, no.~1, pp. 1--16, 2021.

\bibitem{Scutari2014}
G.~Scutari, F.~Facchinei, J.-S. Pang, and D.~P. Palomar, ``Real and {C}omplex
  {M}onotone {C}ommunication {G}ames,'' \emph{IEEE Transactions on Information
  Theory}, vol.~60, no.~7, pp. 4197--4231, 2014.

\end{thebibliography}

\end{document}